\documentclass{aa}
\usepackage{graphicx}
\begin{document}
   \title{The calibration of interferometric visibilities obtained 
   with single-mode optical interferometers}
   \subtitle{Computation of error bars and correlations}

   \author{G. Perrin
          \inst{1}
            }

   \offprints{G. Perrin}

   \institute{LESIA, FRE 2461, Observatoire de Paris, section de 
   Meudon, 5, place Jules Janssen 92190 Meudon, France\\
              \email{guy.perrin@obspm.fr}
              }

  \date{Received 4 September 2002 / Accepted 19 December 2002}

   \abstract{I present in this paper a method to calibrate data 
   obtained from optical and infrared interferometers.  I show that 
   correlated noises and errors need to be taken into account for a 
   very good estimate of individual error bars but also when model 
   fitting the data to derive meaningful model parameters whose 
   accuracies are not overestimated.  It is also shown that under 
   conditions of high correlated noise, faint structures of the source 
   can be detected.  This point is important to define strategies of 
   calibration for difficult programs such as exoplanet detection.  
   The limits of validity of the assumptions on the noise statistics 
   are discussed.   }

   \maketitle
%

\section{Introduction}

   With optical-infrared interferometry becoming more mature, the 
   quality of visibility measurements have become an issue.  
   Single-mode interferometers (see Sect.~2.3) allow one to eliminate 
   non-stationary effects by filtering out the spatial modes of 
   turbulence.  The response of interferometers is therefore very 
   stable and the issue of estimating the accuracies of non-biased 
   data is raised.  The final visibility estimate is a complex 
   quantity as it is a non-linear mix of noisy measurements and of 
   parameter estimates with their own uncertainties.  Estimating the 
   stability of the instrument, a crucial point for calibration, and 
   the final error on visibilities is therefore non-trivial and must 
   be considered with caution.  Moreover, data analysis mainly 
   consists of model fitting the final visibilities and the matter of 
   their potential correlations becomes important, especially if some 
   very faint structures are looked for, as is the case in extra-solar 
   planet detection.  \\
   In this paper I propose a method to meet these challenges.  The 
   method has been tested and elaborated along with the FLUOR 
   interferometer, the first single-mode interferometer.  This method 
   was first published in \cite{perrin1996} and used in 
   \cite{perrin1998}.  It is updated and improved in this paper by 
   accounting for correlations.

\section{Principles of data reduction and calibration}
\begin{figure*}
    \begin{center}
	\hbox{
	      \hspace{-1.4cm}\includegraphics[width=7.7cm]{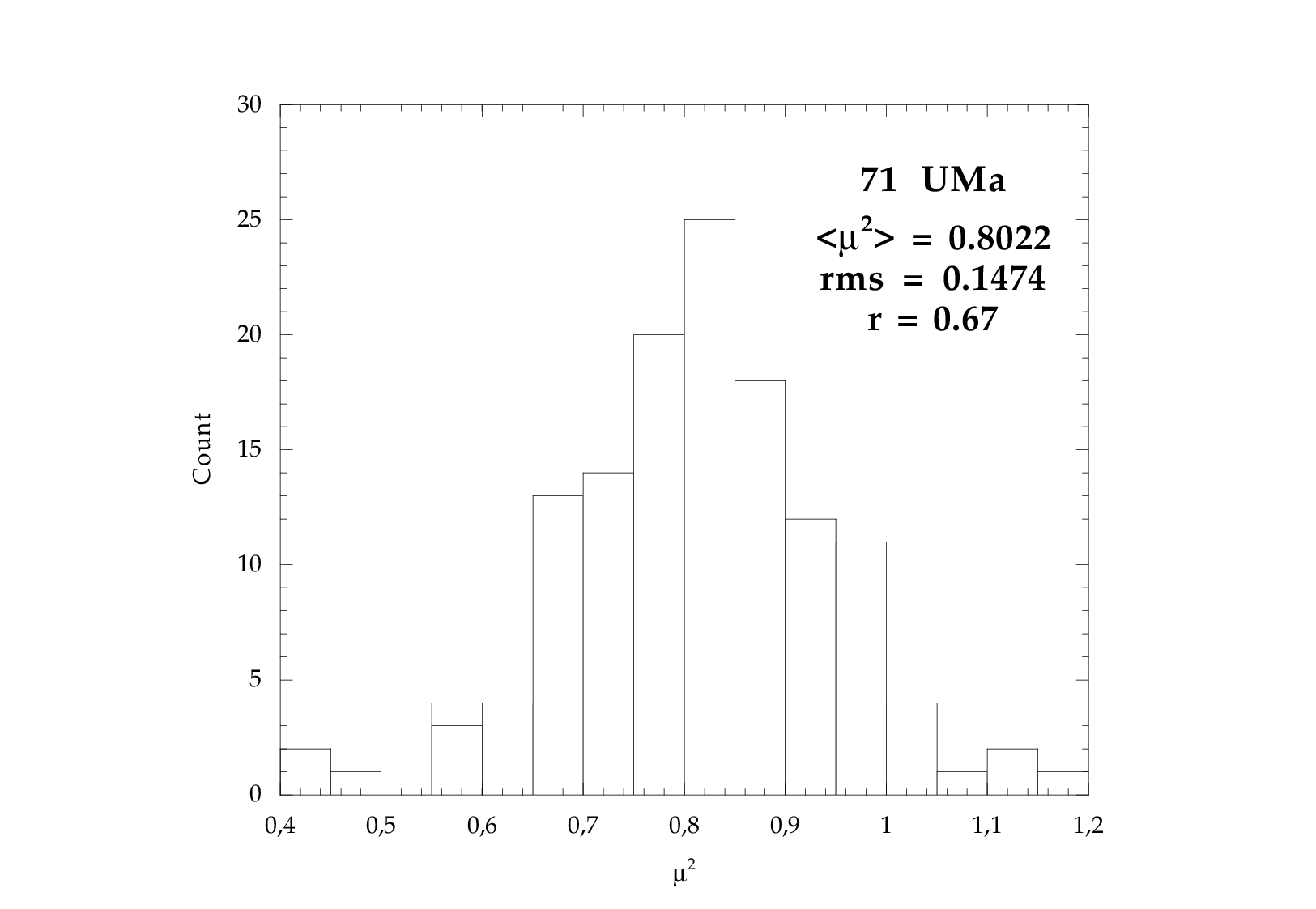}
              \hspace{-1.4cm}\includegraphics[width=7.7cm]{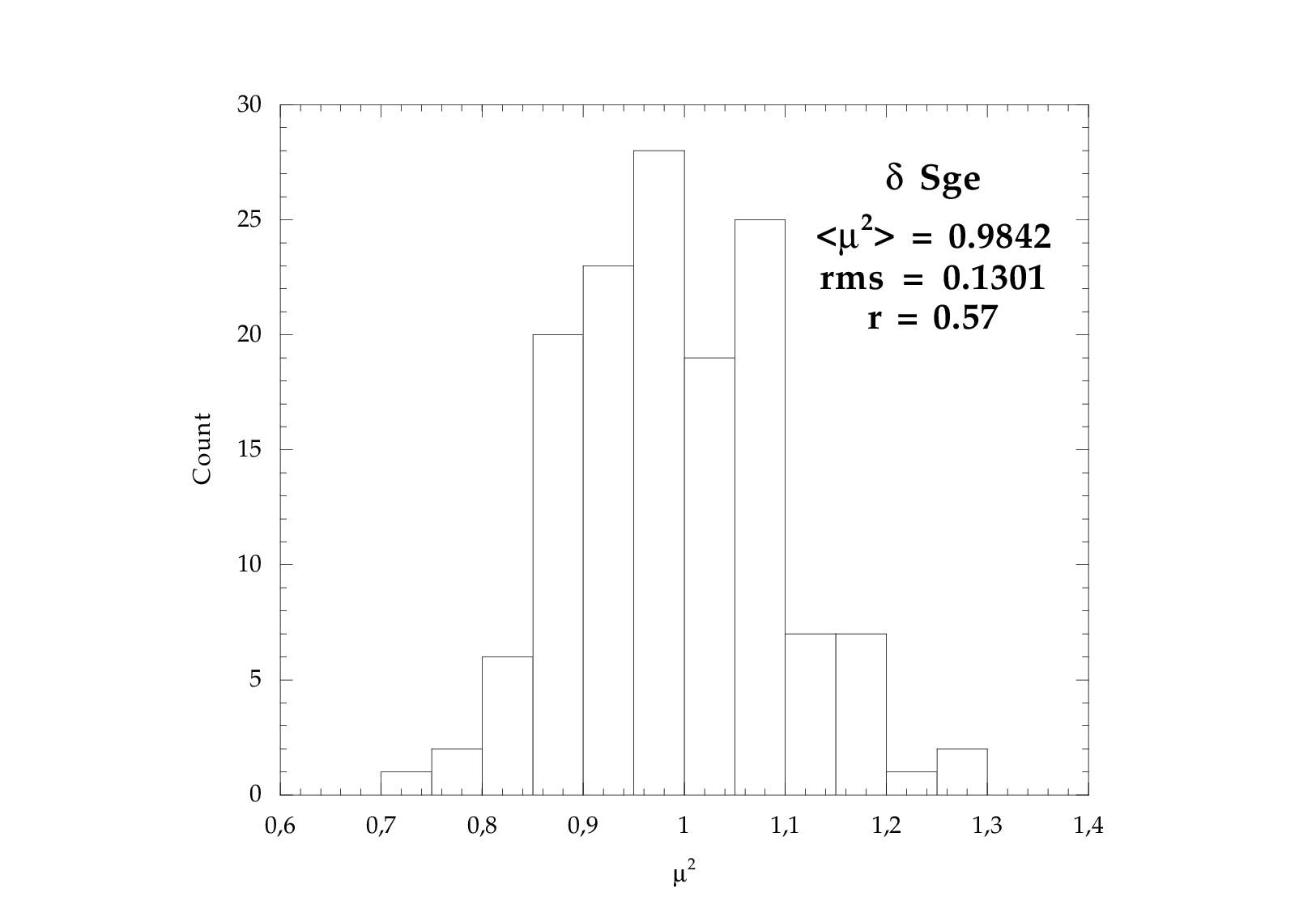}
	      \hspace{-1.4cm}\includegraphics[width=7.7cm]{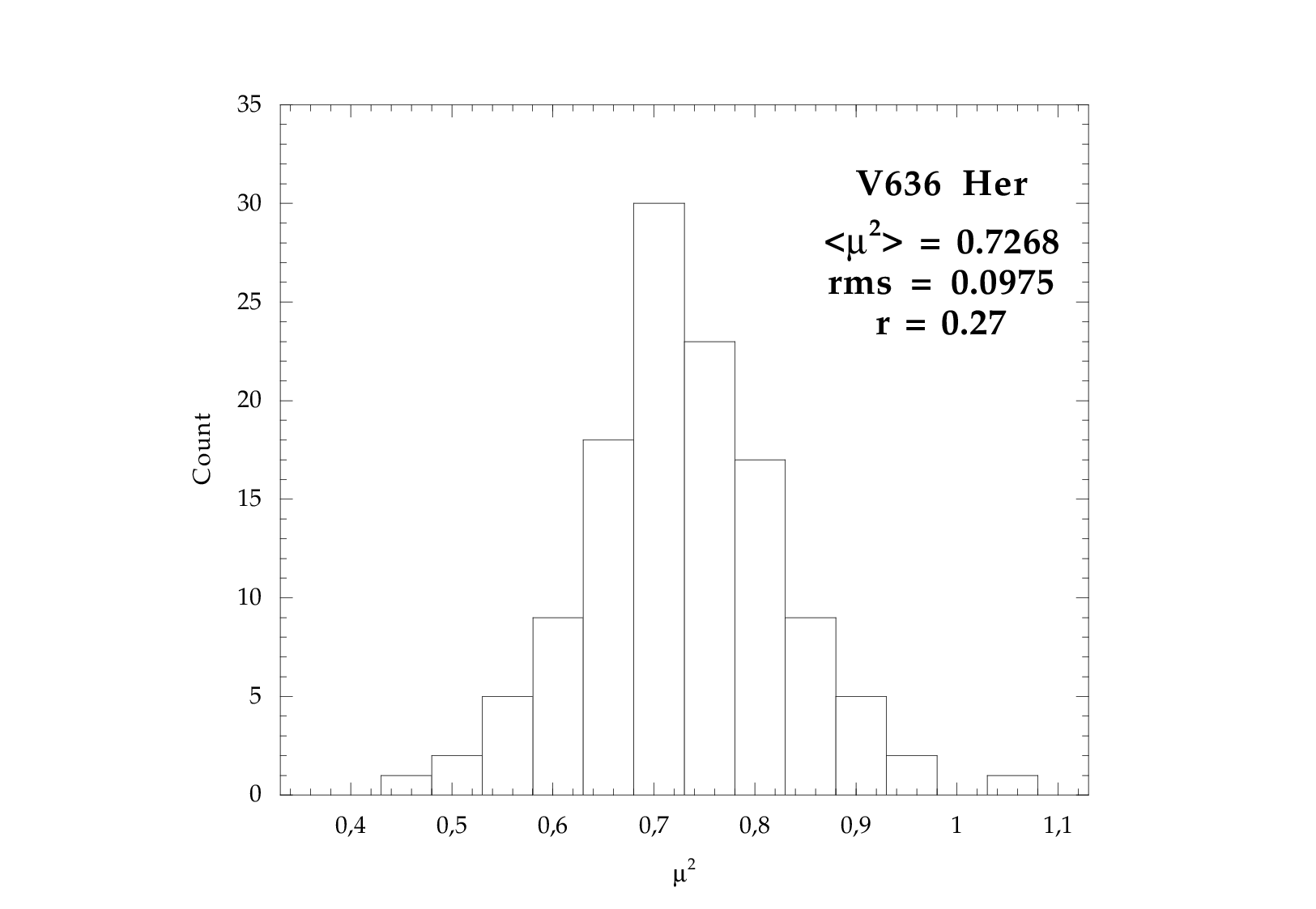}
	}

      \caption{Examples of squared coherence factor histograms 
      obtained with FLUOR in one of the interferometric channels.  
      About 100 interferograms have been recorded for each object. The 
      mean and rms of individual measurements are given for this 
      channel. The correlation factor $r$ measures the noise 
      correlation between the two interferometric channels. The 
      amount of atmospheric piston is decreasing from the left to the 
      right.}
         \label{fig:histo}
   \end{center}
     \end{figure*}
 In this section the general scheme of data reduction is reviewed to 
 introduce the vocabulary and notations.  Two main steps are to be 
 considered.  In the first one (Sect.~2.1), the fringe processing, 
 fringe contrasts are derived from raw signals.  Because of contrast 
 losses, fringe contrasts are calibrated in a second step (Sect.  2.2) 
 to provide the visibilities directly linked to the spatial intensity 
 distribution of the source.

\subsection{Fringe contrasts estimates}
In the following we will distinguish between the fringe contrast 
obtained on a source and the visibility of this source.  The fringe 
contrast measured from a single exposure or scan is called the 
coherence factor and is noted $\mu$ whereas the visibility is noted 
$V$.  \\
Whatever the beamcombining technique, $\mu$ being the modulus of a 
complex number, unbiased estimators are only obtained for squared 
quantities from wich biases due to additive noise can be subtracted.  
In the future, phase referencing techniques may allow one to directly 
measure complex visibilities (real and imaginary parts) but this is 
not the case yet and I will only consider measurements of fringe 
contrast moduli.  The result of the processing of a series of scans on 
a single source is a series of realizations of the $\mu^{2}$ estimator 
or is an average value of the realizations with a 1$\sigma$ error bar 
if their statistical distribution can be trusted to be Gaussian.

\subsection{Necessity for a calibration} 

The average $\mu^{2}$ is not directly an estimator of the squared 
modulus of the visibility of the source because some physical 
phenomena degrade the coherence factor.  Among these phenomena, 
polarization mismatches between the interferometric beams are the most 
common.  Without perfect beam cleaning by a fiber, atmospheric 
turbulence also degrades the fringe contrast.  It is necessary to 
estimate the loss of coherence on a calibrator source for which the 
visibility is known.  A transfer function $T$ is obtained by computing 
the ratio of the measured coherence factor $\mu_{c}$ to the 
expected visibility $V_{exp}$.  With squared quantities this 
yields:
\begin{equation}
    T^{2}=\frac{\mu^{2}_{c}}{V^{2}_{exp}}.
    \label{eq:def_T}
\end{equation}
If the instrument is stable enough then the estimate of $T^{2}$ 
obtained on the calibrator can be used to derive an estimated 
visibility for the source from the measured coherence factor:
\begin{equation}
    V^{2}=\frac{\mu^{2}}{T^{2}}=coT^{2}\,\mu^{2}.
    \label{eq:def_V}
\end{equation}
where I call $coT^{2}$ the squared co-transfer function. Its use will 
be detailed in Sect.~4. 
\subsection{Assumptions in the case of a single-mode interferometer}
\begin{figure*}[t]
    \begin{center}
	\vbox{
	\includegraphics[width=15cm]{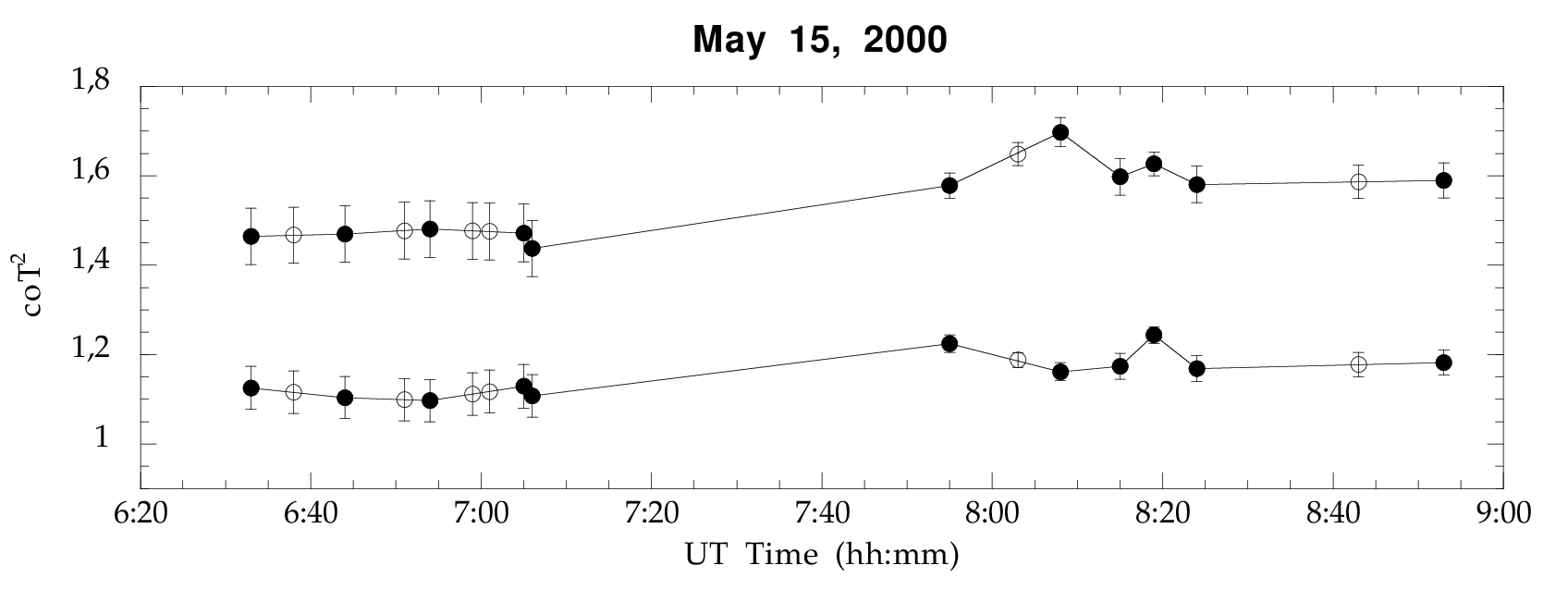}
	\includegraphics[width=15cm]{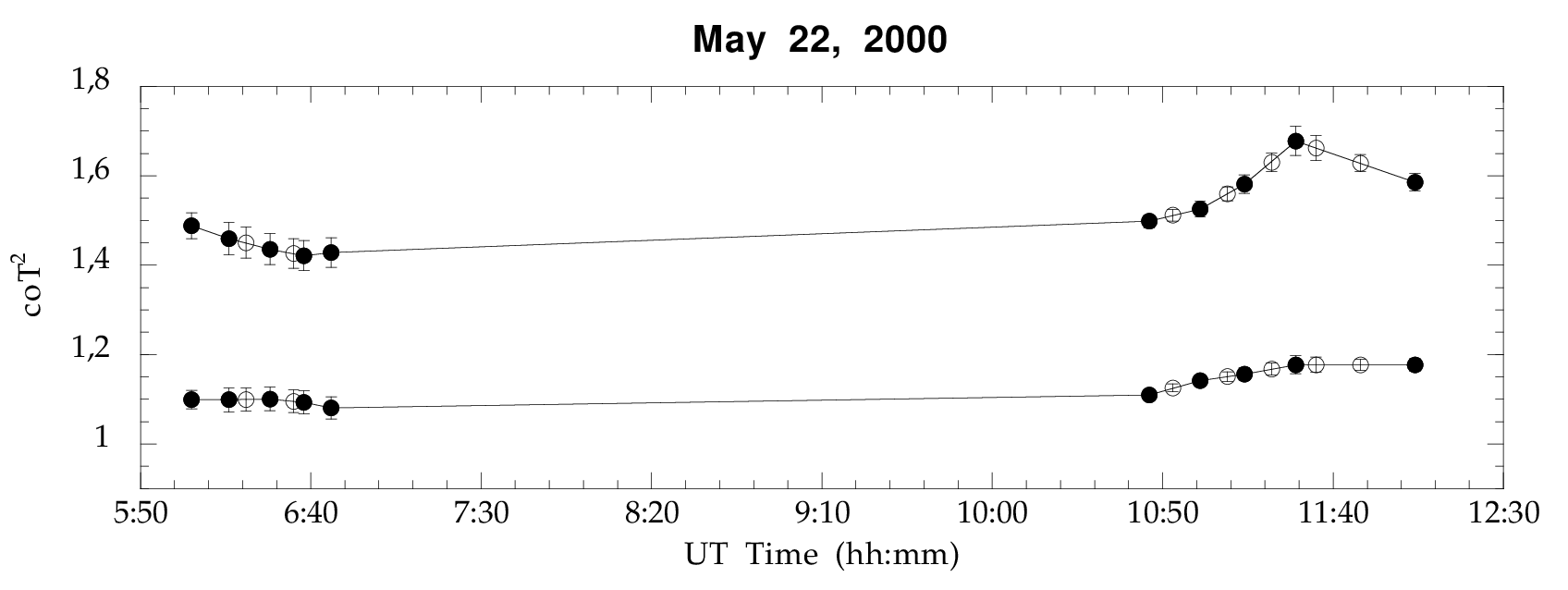}
	     }
      \caption{Examples of squared co-transfer functions measured with 
      FLUOR. The two curves for each night correspond to the two 
      interferometric channels of the coaxial interferometer. The 
      full circles are the squared co-transfer functions measured on 
      calibrators whereas the open circles are the values 
      interpolated at the time when the science targets were observed. 
      $1\,\sigma$ error bars are displayed.}
         \label{fig:coT2}
   \end{center}
     \end{figure*}
\begin{figure*}
    \begin{center}
	\includegraphics[width=18cm]{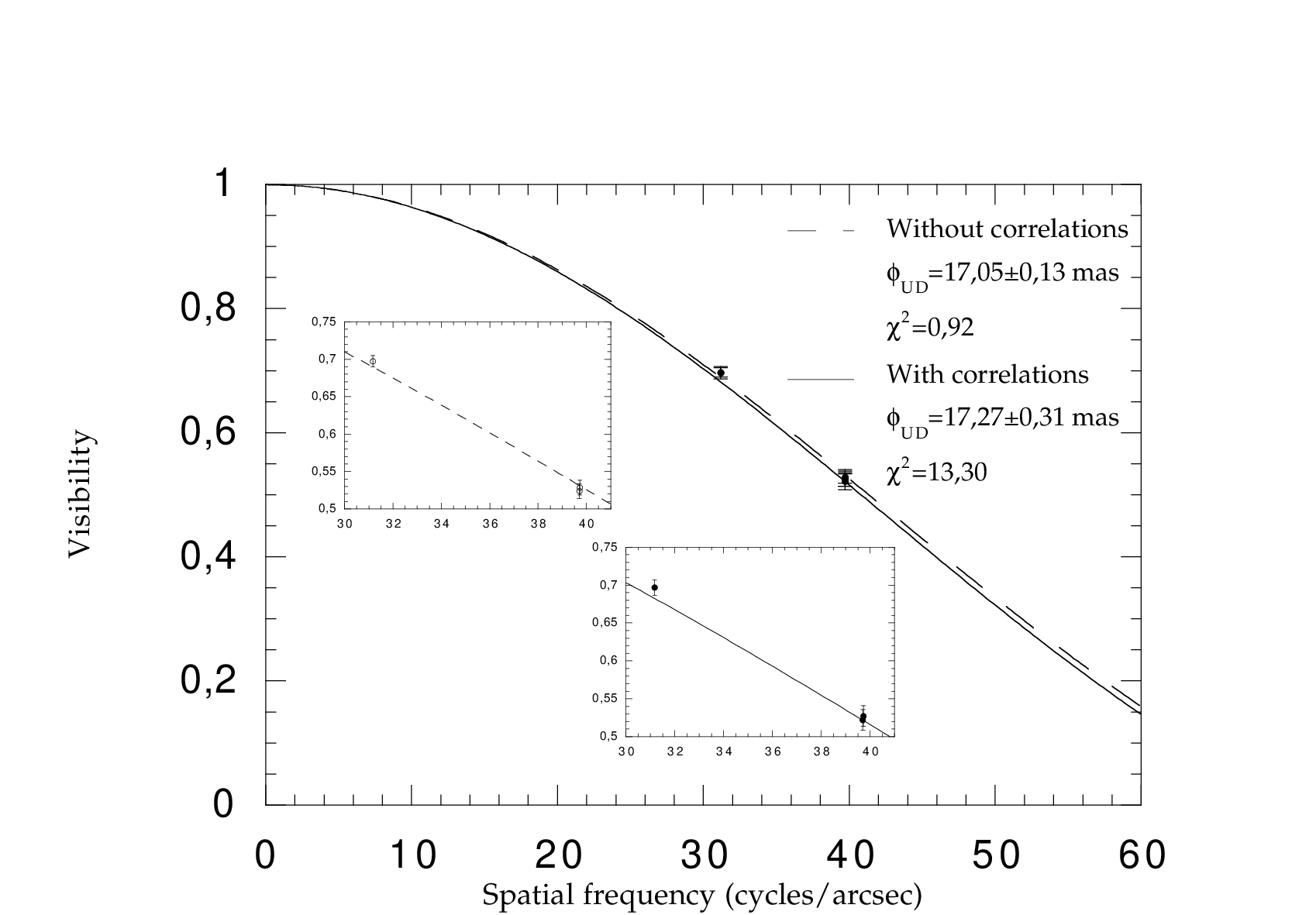}
       \caption{Example of visibility fit.  The source is SW~Vir and 
       the model is a uniform disk.  Errors are $1\,\sigma$ errors.  
       Open circles and dashed line are the visibilities and model fit 
       computed without taking correlations into account.  Full 
       circles and continuous line are the equivalent with the method 
       described in this paper.  The two cases are separately 
       presented in the little windows.  }
         \label{fig:swvir}
   \end{center}
     \end{figure*}
\begin{figure*}
    \begin{center}
	\vbox{
	\hbox{
		\includegraphics[width=9cm]{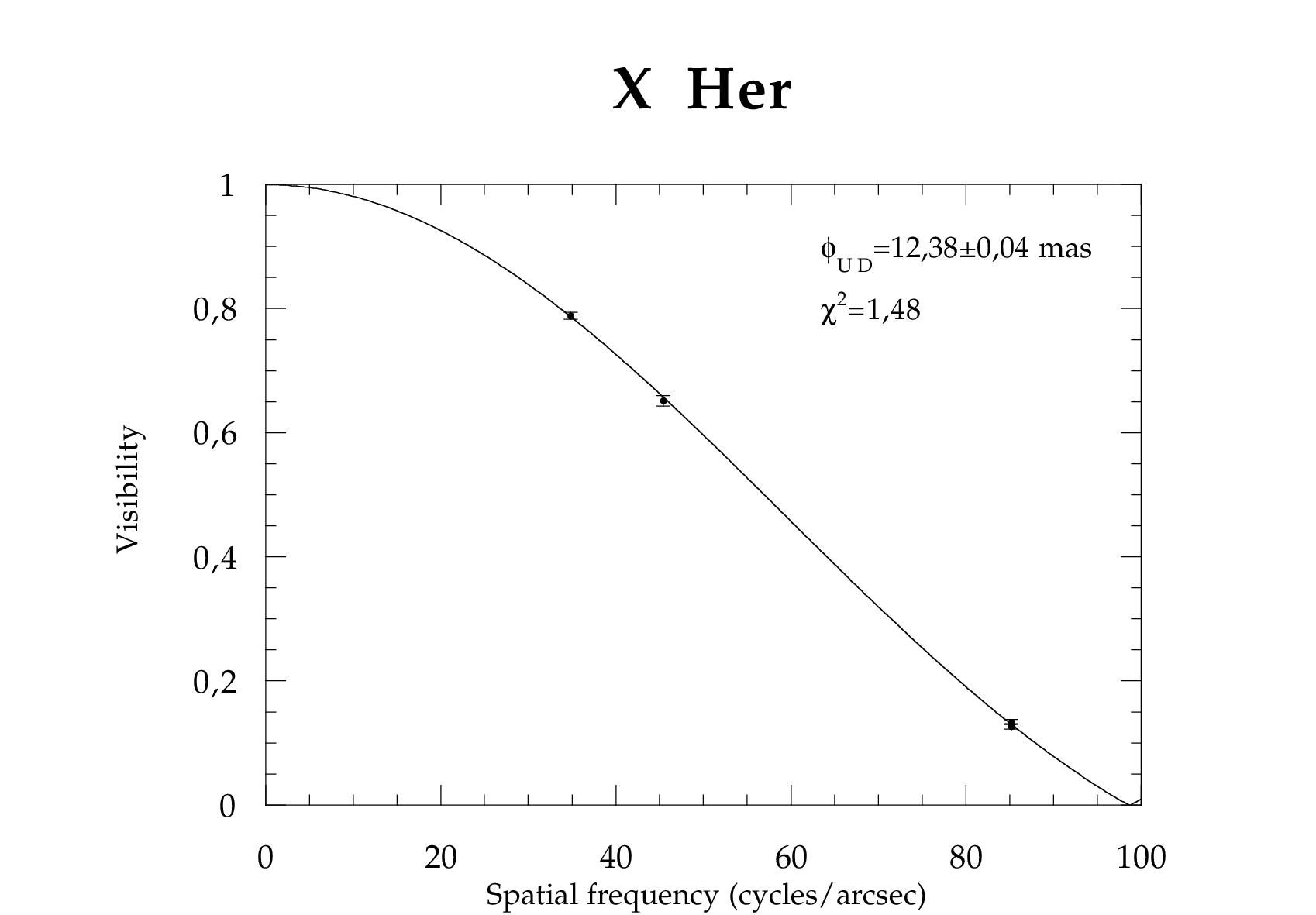}
		\hspace{-1cm}
		\includegraphics[width=9cm]{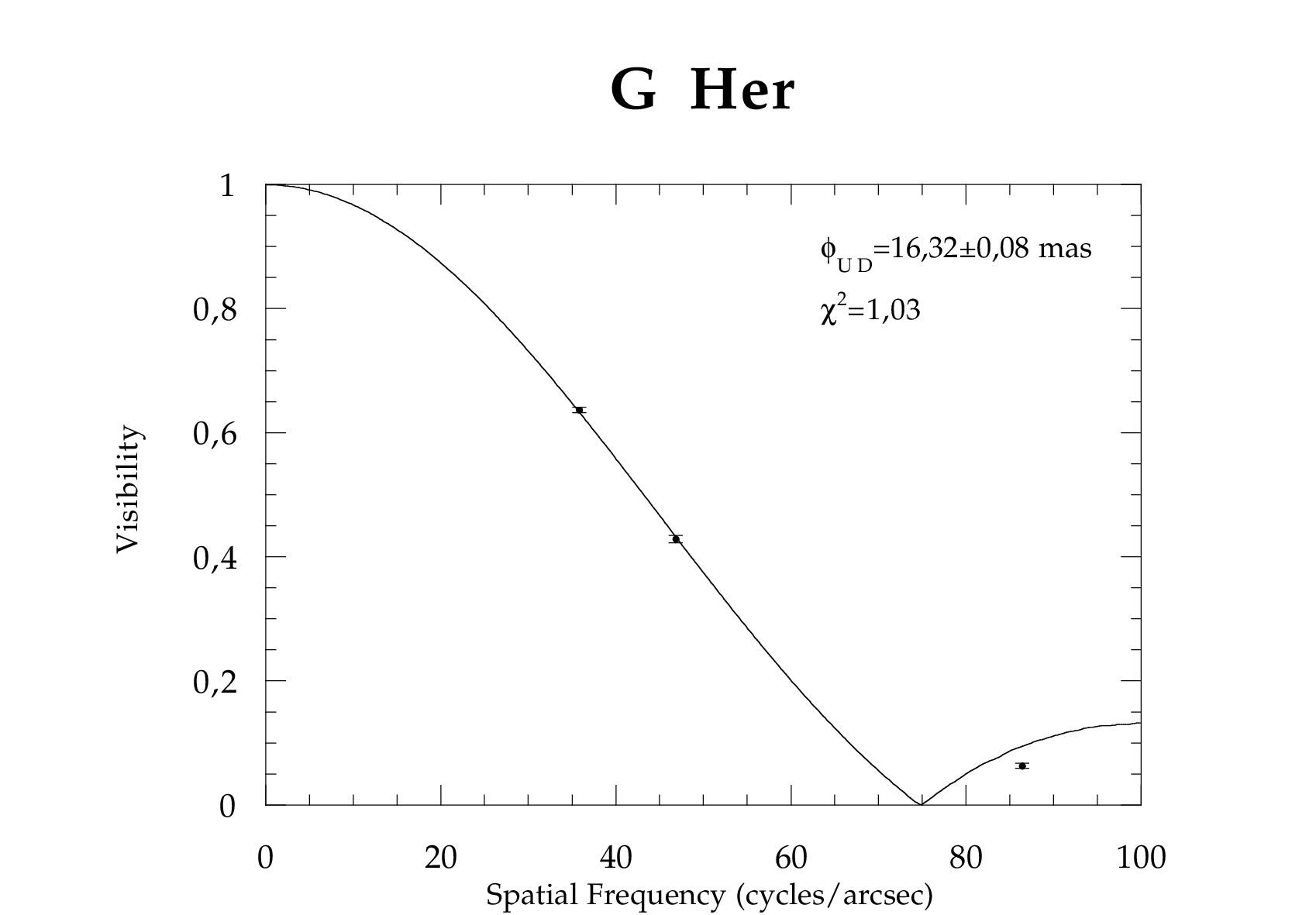}
	}
	\hbox{
		\includegraphics[width=9cm]{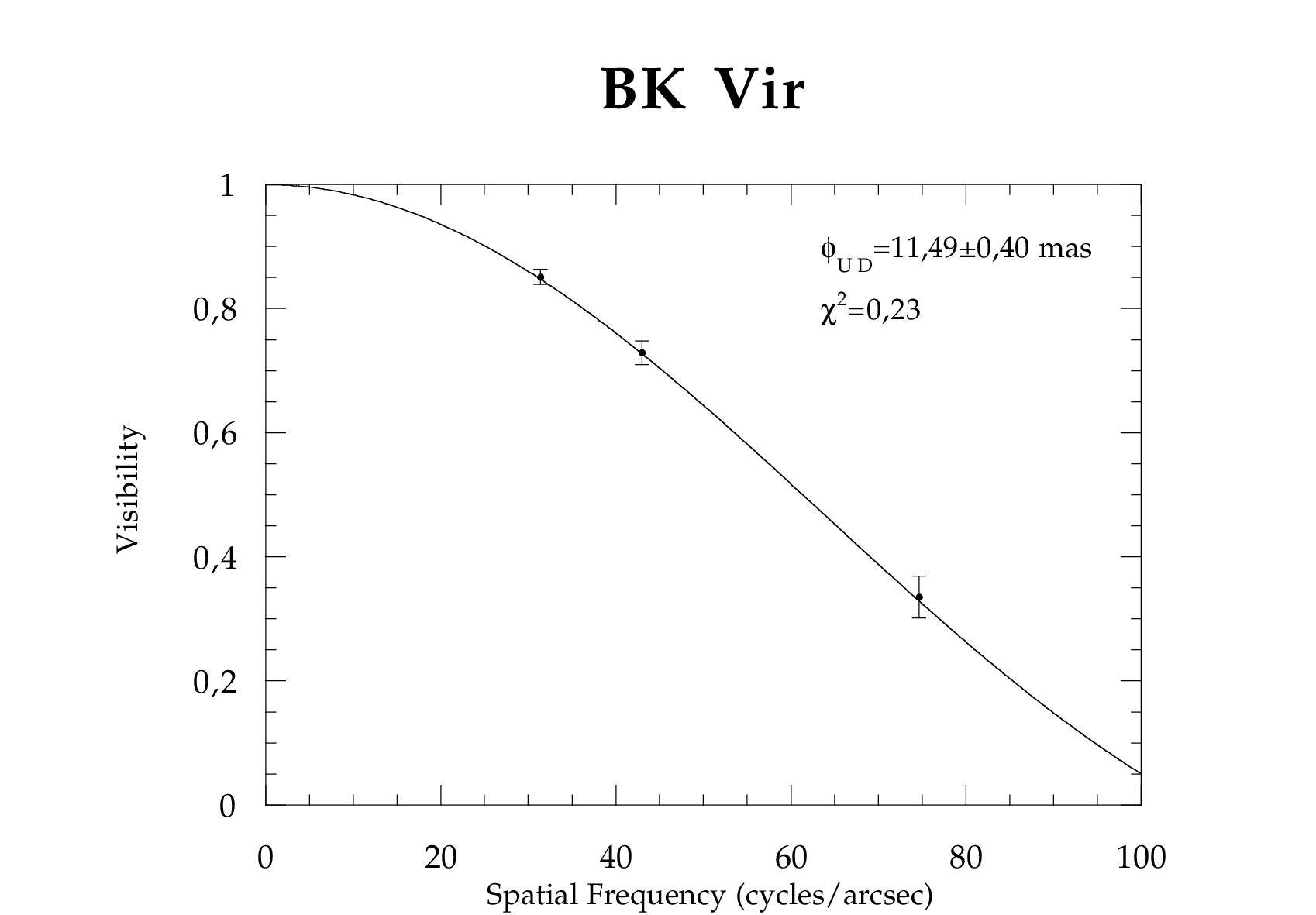}
		\hspace{-1cm}
		\includegraphics[width=9cm]{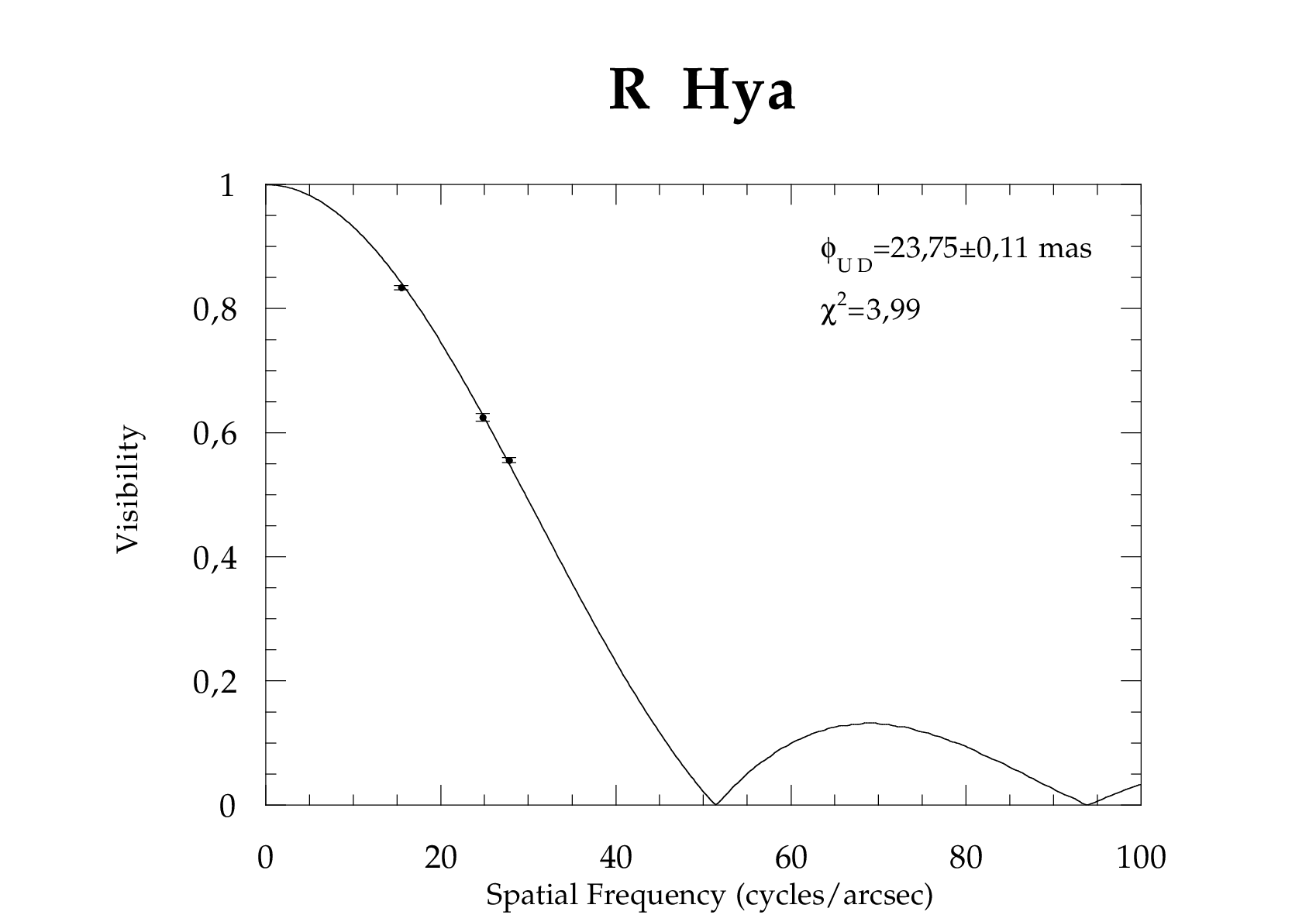}
	} } 
	\caption{Examples of fits of visibility data with a 
	uniform disk model.  All correlations are taken into account.  
	Errors are $1\,\sigma$ errors.  The visibility point 
	around 85\,~cycles/arcsec in the G~Her panel is not taken into 
	account in the fit.}
         \label{fig:exemples}
   \end{center}
     \end{figure*}
The calibration process may fail if the assumption that the transfer 
function is stable is wrong.  This usually happens if non-stationary 
processes like atmospheric phase turbulence play a role in the fringe 
formation process.  In a perfect single-mode interferometer in which 
single-mode fibers are used to filter the phase aberrations produced 
by atmospheric turbulence (except for the piston mode) these 
non-stationary effects are eliminated.  In order to avoid 
instabilities due to the piston mode, the fringes are scanned at a 
frequency far above the characteristic frequency of piston.  In 
interferometers where the piston is stabilized by a fringe tracking 
servo loop, this issue is solved.  The remaining main sources of 
variation of the transfer function are basically temperature drifts 
and differential polarisation effects due to the change of beam 
inclination on the first mirrors with changing positions of the 
sources in the sky.  In both cases the transfer function drifts are 
very slow and a good estimate of the transfer function can be obtained 
by interpolating two estimates bracketing the source to be calibrated.  
This has been demonstrated with the FLUOR beamcombiner, as will be 
shown is Sect.~7.2.  \\
In the following we will therefore consider that the efficiency of the 
interferometer is continuously assessed by observing calibrators 
before and after science sources.  We will not consider the case where 
the transfer function is derived by averaging individual transfer 
functions on a large temporal scale, as this is not required for a 
single-mode interferometer.  This technique does not allow one to 
assess the quality of the calibration in detail.

\section{Estimating fringe contrasts}
This section focuses on estimating the statistical properties of 
fringe contrasts.  I will not describe the method to compute coherence 
factors from single exposures and I will refer the reader to 
appropriate articles in the next paragraphs.

\subsection{Single-channel spatial modulation interferometer} 
In a multiaxial interferometer, distant parallel beams feed a focusing 
optic.  The beams are recombined in the focal plane where they overlap 
at the focus locus.  The modulation is spatial as the fringe phase 
varies across the diffraction pattern.  A method to derive fringe 
contrasts has been published by \cite{mourard1994} in the case of 
GI2T. The method has been adapted to AMBER which is a single-mode 
multiaxial interferometer (Chelli {\it et al.}~2000).\\
Thanks to the filtering of the non-stationary modes of turbulence, the 
statistics of $\mu^{2}$ can be well approximated by a Gaussian 
distribution.  This will be demonstrated in the case of the data 
obtained with FLUOR in Sect.~7.1.  The estimate of the squared 
coherence factor is therefore the mean of the distribution of the 
realizations denoted $\overline{\mu^{2}}$.  An unbiased estimate of 
the variance of individual measurements is:
\begin{equation}    
S^{2}=\frac{1}{N-1}\sum_{n=0}^{N-1}(\mu^{2}_{n}-\overline{\mu^{2}}).
\end{equation}
The estimate of the variance of the coherence factor estimator 
$\overline{\mu^{2}}$ is then:
\begin{equation}
    Var(\overline{\mu^{2}})=\frac{S^{2}}{N}.
\end{equation}
\subsection{Two-channel temporal modulation interferometer}
In a coaxial interferometer, beams are superimposed in position and in 
direction.  This can be realized with a beamsplitter or with a fiber 
coupler.  A relative phase between the beams is introduced by setting 
an optical path difference.  This is achieved with a moving mirror in 
one of the two beams, hence the temporal modulation of the phase.  A 
method to compute fringe contrasts for this type of interferometer is 
described in \cite{foresto1997}.  A more recent method based on 
wavelets analysis has been proposed by \cite{segransan2002}.  The 
method to obtain an estimate of the coherence factor without the 
photon noise bias is explained in \cite{perrin2002}.  A prototype 
instrument for this kind of interferometer is the FLUOR beamcombiner.  
\\
The difference with the previous interferometer of Sect.~3.1 is that 
it produces two interferometric beams and therefore two sets of 
coherence factors estimates.  The statistics of each set can be well 
approximated by a gaussian statistics as will be shown in 
Sect.~\ref{sec:examples}. The photon noises of the two 
interferometric signals are uncorrelated.  The read-out noises are 
generally considered uncorrelated but some correlation may occur as 
different pixels share the same read-out electronics.  In addition the 
two beams suffer from the same turbulence effects (residual piston and 
photometric beam fluctuations) which generate some noise in the 
measurements.  Part of the noise is therefore common to the two 
signals and the coherence factors estimates are correlated.  The 
correlation factor $r$ is directly estimated from the $\mu^{2}$ 
distributions:
\begin{equation}
r=\frac{\langle\mu^{2}_{1}-\overline{\mu^{2}_{1}}\rangle\langle\mu^{2}_{2}-\overline{\mu^{2}_{2}}\rangle}{\sqrt{Var(\mu^{2}_{1})Var(\mu^{2}_{2})}},
\end{equation}
where the subscripts describe the two interferometric channels.

\section{Estimating the transfer function}
It is assumed that the transfer function is a slowly varying function 
which is rapidly sampled.  This property will be illustrated with real 
data in Sect.~\ref{sec:examples}.  It is then legitimate to linearly 
interpolate the squared transfer function at the time when the science 
source was observed.  Because the variances of products of random 
variables are more easily calculated than those of ratios, the 
reciprocal of the squared transfer function, the squared co-transfer 
function, is interpolated instead of the squared transfer function.  
The use of one or the other is equivalent.  In order to be general, 
two interferometric outputs are always considered.  The particular 
case of the multiaxial interferometer will be considered in 
discussions.  The expression of the interpolated co-transfer functions 
in the two outputs of the instrument is :
\begin{equation}
   \left\{\begin{array}{ccc}
       coT_{1}^{2} & = & 
   x\frac{{V^{a}}^{2}}{\overline{{\mu_{1}^{a}}^{2}}}+(1-x)\frac{{V^{b}}^{2}}{\overline{{\mu_{1}^{b}}^{2}}}\\    
   coT_{2}^{2} & = & 
   x\frac{{V^{a}}^{2}}{\overline{{\mu_{2}^{a}}^{2}}}+(1-x)\frac{{V^{b}}^{2}}{\overline{{\mu_{2}^{b}}^{2}}}
      \end{array}\right.
    \label{eq:cot2}
\end{equation}
$V^{a}$ and $V^{b}$ are the expected visibilities of calibrators A and 
B. Coefficient $x$ is the relative time distance between the 
observation of the science target and the observation of calibrator A. 
The expected visibilities are supposed to be Gaussian random 
variables.  Dropping the channel indices, the variances of the squared 
co-transfer functions are equal to:
\begin{equation}
    Var(coT^{2})=x^{2}Var({coT^{a}}^{2})+(1-x)^{2}Var({coT^{b}}^{2})
\end{equation}
when the two calibrators are different. When the two calibrators are 
the same then the variance is equal to:
\begin{eqnarray}
Var(coT^{2}) = & & x^{2}Var({coT^{a}}^{2})+(1-x)^{2}Var({coT^{b}}^{2}) 
\\ \nonumber & & +2\frac{x(1-x)}{ 
\overline{{\mu^{a}}^{2}}\overline{{\mu^{b}}^{2}}}Var({V^{a}}^{2})
\end{eqnarray}
The squared co-transfer functions estimated on the calibrators are 
ratios of Gaussian estimators.  These new random distributions are not 
Gaussian.  They are Cauchy distributions (the density probability of 
which is a Lorentzian) with no mean and no variance.  By analogy with 
the standard deviation of a Gaussian law, an estimate of the 
uncertainty can be derived from the width of the confidence interval.  
The upper and lower limits of the confidence interval at 68.3\% for a 
ratio of two Gaussian random variables $\alpha$ and $\beta$ are given 
by (\cite{pelat1992}):
\begin{equation}
L_{\pm}=\frac{ 
\alpha \pm \frac{\beta / \sigma_{\beta}} {\left( \left( \alpha / 
\sigma_{\alpha}\right)^{2}+ \left( \beta / \sigma_{\beta} 
\right)^{2}-1\right)^{1/2} }\,\sigma_{\alpha} } { \beta \mp 
\frac{\alpha / \sigma_{\alpha}} {\left( \left( \alpha / 
\sigma_{\alpha}\right)^{2}+ \left( \beta / \sigma_{\beta} 
\right)^{2}-1\right)^{1/2} }\,\sigma_{\beta} }
\label{eq:erreur}
\end{equation}
This interval is not symmetric.  I choose as error bar the larger 
distance of the mean to the limits, thus slightly overestimating the 
error.  The probability that the true value is in this interval around 
the mean is therefore larger than 68.3\%.  In the following I will 
consider that ratios of Gaussian variables are Gaussian variables.  
This is not rigourously true but this allows one to derive expressions 
otherwise difficult to handle.  The Gaussian and Lorentzian functions 
are both bell curves, the wings of the latter being more extended than 
that of the Gaussian.  The approximation amounts to giving more weight 
to the center of the lorentzian.  The other possibility would be 
to apply Monte-Carlo or bootstrap methods to all error bar 
computations, which would make data reduction a very long process for a 
very limited gain. Nevertheless, this method will have to be applied 
once to compute the correlation between the visibilities of a 
two-channel coaxial interferometer (Sect.~\ref{sec:vis}).  Consistency 
of error bars will be addressed in Sect.~\ref{sec:examples}.  This 
final consistency is the justification of the approximations 
performed.

\section{Estimating visibilities}
\label{sec:vis}
\subsection{Mean and variance of channel visibilities}
The single-channel squared visibility in the case of a multiaxial 
interferometer or the two-channel squared visibilities of a coaxial 
interferometer are simply the product of the science target squared 
coherence 
factors and squared co-transfer functions of Eq.~(\ref{eq:cot2}) 
yielding:
\begin{equation}
    \label{eq:vis}
\left\{\begin{array}{ccc}
    V_{1}^{2} & = & 
x\frac{\overline{\mu_{1}^{2}}}{\overline{{\mu_{1}^{a}}^{2}}}{V^{a}}^{2}+(1-x)\frac{\overline{\mu_{1}^{2}}}{\overline{{\mu_{1}^{b}}^{2}}}{V^{b}}^{2}\\    
V_{2}^{2} & = & 
x\frac{\overline{\mu_{2}^{2}}}{\overline{{\mu_{2}^{a}}^{2}}}{V^{a}}^{2}+(1-x)\frac{\overline{\mu_{2}^{2}}}{\overline{{\mu_{2}^{b}}^{2}}}{V^{b}}^{2}
\end{array}\right.
\end{equation}
The variances of the two-channel squared visibilities are calculated 
with the variances of the squared co-transfer functions and of the 
squared coherence factors with the following formula :
\begin{eqnarray}
    Var(AB)= & & Var(A)Var(B)+\bar{A}^{2}Var(B)\\ \nonumber
             & & +\bar{B}^{2}Var(A)
\label{eq:varprod}
\end{eqnarray}
We therefore have:
\begin{eqnarray}
    \label{eq:varvis}
    Var(V_{1}^{2}) = & & Var(\overline{\mu_{1}^{2}})Var(coT_{1}^{2})+ 
    \overline{\mu_{1}^{2}}Var(coT_{1}^{2})\\ \nonumber 
                     & & +coT_{1}^{2}Var(\overline{\mu_{1}^{2}})\\ \nonumber    
    Var(V_{2}^{2}) = & & Var(\overline{\mu_{2}^{2}})Var(coT_{2}^{2})+ 
    \overline{\mu_{2}^{2}}Var(coT_{2}^{2})\\ \nonumber 
                     & & +coT_{2}^{2}Var(\overline{\mu_{2}^{2}})
\end{eqnarray}
\subsection{Correlation between two channel visibilities}
\label{sec:2chan}
The above equations define the uncertainties on the channel estimates 
of the visibilities.  In the case of a multiaxial interferometer, this 
is the final estimate of the visibility.  In the case of coaxial 
interferometers, the two estimates of the visibility need to be 
averaged at this stage.  For this, it is necessary to assess the 
correlation factor between the two estimates.  By definition, the 
correlation factor is equal to:
\begin{equation}
\rho_{12}=\frac{\langle(V_{1}^{2}-\overline{V_{1}^{2}})(V_{2}^{2}-\overline{V_{2}^{2}})\rangle}{\sqrt{Var(V_{1}^{2})Var(V_{2}^{2})}}
\label{eq:rho12}
\end{equation}
This quantity is defined by sums, ratios and products of ten random 
variables.  The correlation factor has to be computed with a 
Monte-Carlo method by simulating each random variable from its mean 
and variance (assuming it has Gaussian statistics) and by correlating 
the series of $V_{1}^{2}$ and $V_{2}^{2}$.  In the special case when 
the correlations between measured quantities can be neglected because 
the correlated noise level is far below that of the uncorrelated 
noise, it is to be noticed that there is still a correlation due to 
the common estimated values of the calibrators expected visibilities 
in the two interferometric channels.  This can be illustrated by the 
equation below when the first and second calibrators are different:
\begin{equation}
    \rho_{12}=
    \frac{
          x^{2} \frac{\overline{
                                \mu_{1}^{2}
                               }
	             }
                     {\overline{
                                {\mu_{1}^{a}}^{2}}
			       }
                \frac{\overline{
		                \mu_{2}^{2}
			       }
		     }
		     {\overline{
		                {\mu_{2}^{a}}^{2}}
				}
          Var({V^{a}}^{2})+(1-x)^{2}
	        \frac{\overline{
		                \mu_{1}^{2}
			       }
	             }
		     {\overline{
		                {\mu_{1}^{b}}^{2}}
			       }
	        \frac{\overline{
		                \mu_{2}^{2}
			       }
		     }
		     {\overline{
		                {\mu_{2}^{b}}^{2}}
				}
	  Var({V^{b}}^{2})
	 }
         {
	  \sqrt{Var(V_{1}^{2})Var(V_{2}^{2})}
         }
\end{equation}
When the two calibrators are the same the correlation factor has the 
particular expression (for sake of simplicity the two estimates of 
the visibility at slighlty different baselines are supposed to be the 
same):
\begin{equation}
    \rho_{12}\!=\!
    \frac{
          \left[
	        x\frac{\overline{
		                 \mu_{1}^{2}
				}
		      }
		      {\overline{
		                 {\mu_{1}^{a}}^{2}
				}
		      }\!+\!
	        (1-x)
		 \frac{\overline{
		                 \mu_{1}^{2}
				}
		      }
		      {\overline{
		                 {\mu_{1}^{b}}^{2}
				}
		      }
	  \right]\!\!\!
\left[
	        x\frac{\overline{
		                 \mu_{2}^{2}
				}
		      }
		      {\overline{
		                 {\mu_{2}^{a}}^{2}
				}
		      }\!+\!
	        (1-x)
		 \frac{\overline{
		                 \mu_{2}^{2}
				}
		      }
		      {\overline{
		                 {\mu_{2}^{b}}^{2}
				}
		      }
	  \right]\!\!\!	  Var({V^{a}}^{2})
	 }
	 {
	  \sqrt{Var(V_{1}^{2})Var(V_{2}^{2})}
	 }
\end{equation}
It is now easy to see from Eq.~(\ref{eq:vis}) that if the noise on the 
measurements is negligible with respect to the uncertainties on the 
expected visibilities of the calibrators then the correlation tends 
towards 1.  In this case, the second interferometric channel brings no 
extra information except a consistency check and the precision on the 
visibility of the science target is directly proportional to the 
precision on the expected visibilities of the calibrators.

\subsection{Comments on visibility variances}
It is also interesting to analyze the propagation of noises in the 
visibility estimates.  For example, if the noise on the measurements 
is negligible, it is possible to evaluate the amount of variance due 
to the uncertainty on the calibrators visibilities (or diameters).  
Dropping channel indices I obtain:
\begin{eqnarray}
Var(V^{2})|_{\mu,\mu^{a},\mu^{b}}= & & 
x^{2}\left[\frac{\overline{\mu^{2}}}{
\overline{{\mu^{a}}^{2}}}\right]^{2}Var({V^{a}}^{2})\\ \nonumber
& & 
+(1-x)^{2}\left[
\frac{\overline{\mu^{2}}}
{\overline{{\mu^{b}}^{2}}}
\right]^{2}Var({V^{b}}^{2})
\end{eqnarray}
If the uncertainties on the calibrators\'~squared visibilities are 
equal to 1\% and if the coherence factors are all equal to 1 and 
observations are equally spaced in time then the uncertainty on the 
measured squared visibility is equal to 0.7\%.  This equation also 
shows that the smaller the visibility of the calibrator, the more 
amplified the noise on its expected visibility is.  Symmetrically, the 
smaller the visibility of the science target, the smaller the 
contribution to the noise of the calibrators\'~expected visibilities.


\subsection{Final estimate of the visibility in a two-channel 
interferometer}

The final squared visibility $V^{2}$ is estimated from the two 
squared 
visibilities obtained from each output of the interferometer 
$V_{1}^{2}$ and $V_{2}^{2}$ and their respective variances (or 
equivalently uncertainties $\sigma_{1}$ and $\sigma_{2}$).  I define 
the final estimate $V^{2}$ as being the least squares fit estimator 
of 
the squared visibility as this is an optimal estimator for Gaussian 
random variables.  In this fit, the model is linear and has only one 
parameter: $V^{2}$.  Let us call $C$ the covariance matrix of 
$V_{1}^{2}$ and $V_{2}^{2}$:
\begin{equation}
    C=\left[
    \begin{array}{cc}
        \sigma_{1}^{2} & \rho_{12}\sigma_{1}\sigma_{2}  \\
        \rho_{12}\sigma_{1}\sigma_{2} & \sigma_{2}^{2}
    \end{array} 
      \right].
\end{equation}
The quantity to minimize in the least squares fit can then be written:
\begin{equation}
    S(V^{2})=\left[
    \begin{array}{c}
        V_{1}^{2}-V^{2}  \\
        V_{2}^{2}-V^{2}
    \end{array} 
      \right]^{t}C^{-1} \left[
     \begin{array}{c}
        V_{1}^{2}-V^{2}  \\
        V_{2}^{2}-V^{2}
    \end{array} \right] = Y^{t}C^{-1}Y.
\end{equation}
It can be shown that the minimum is reached for:
\begin{equation}
    V^{2}=(X^{t}C^{-1}X)^{-1}X^{t}C^{-1}\left[
    \begin{array}{c}
        V_{1}^{2}  \\
        V_{2}^{2}
    \end{array} 
      \right].
\end{equation}
with
\begin{equation}
   X=\left[
    \begin{array}{c}
        1  \\
        1
    \end{array} 
      \right],
\end{equation}
the uncertainty on $V^{2}$ being:
\begin{equation}
    \sigma_{V^{2}}^{2}=(X^{t}C^{-1}X)^{-1}X^{t}C^{-1} .
\end{equation}
The above equations yield the final visibility estimate:
\begin{equation}
    \label{eq:visfinale}
V^{2}=\frac{V_{1}^{2}(\sigma_{2}^{2}-\rho_{12}\sigma_{1}\sigma_{2})+V_{1}^{2}(\sigma_{1}^{2}-\rho_{12}\sigma_{1}\sigma_{2})}{\sigma_{1}^{2}\sigma_{2}^{2}-2\rho_{12}\sigma_{1}\sigma_{2}}
\end{equation}
and the associated error:
\begin{equation}
\sigma_{V^{2}}^{2}=\frac{(1-\rho_{12}^{2})\sigma_{1}^{2}\sigma_{2}^{2}}{\sigma_{1}^{2}+\sigma_{2}^{2}-2\rho_{12}\sigma_{1}\sigma_{2}}
\end{equation}
If $\rho_{12}=1$ then the two single-output squared visibilities 
$V_{1}^{2}$ and $V_{2}^{2}$ are fully correlated and the above 
expression does not apply.  In this case V is equal to one of the two 
single-output visibilities with its associated error bar.\\ 
The quality of the fit is expressed by the $\chi^{2}$:
\begin{equation}
    \label{eq:chi2}
\chi^{2}=\frac{(V_{1}^{2}-V_{2}^{2})^{2}}{\sigma_{1}^{2}+\sigma_{2}^{2}-2\rho_{12}\sigma_{1}\sigma_{2}}
\end{equation}
This parameter is important because it allows us to check the 
consistency of the instrument and of the method to measure the 
visibilities and the error bars.  If all assumptions are correct then 
the $\chi^{2}$ should be equal to 1 on average.  In the FLUOR software 
we use this number as a data quality parameter.  Data with $\chi^{2}$ 
greater than 3 should be examined in detail and rejected for science 
programs requiring a very good quality of calibration as the 
probability to get a value larger than 3 is only of 8.33\%.

\subsection{Correlations of multiple baseline interferometer simultaneous visibilities}
Coherence factors recorded simultaneously on different baselines with 
telescopes in common may also be correlated.  This correlation should 
be taken into account and saved with the reduced data in the form of a 
correlation matrix.  The correlations may be as high as the 
correlations between the two channels of a coaxial interferometer as 
all calibrators are common to all baselines.  The method used in 
Sect.~\ref{sec:2chan} should be applied.  A correlation matrix for 
the $\overline{\mu^{2}}$ should be computed first.  The final 
correlation factors for the final visibility estimates are then 
computed with a Monte-Carlo method.

\section{Correlations between  non-simultaneous visibilities}
Visibilities obtained on different baselines or on different days are 
usually considered independent.  In the last paragraph, we focused on 
the possible correlations of visibilities recorded simultaneously on 
baselines with telescopes in common.  In this section, we will 
consider the correlation due to common uncertainties in the 
calibration process for independent baselines or for visibilities 
measured at different times.  The calibration of the transfer function 
may have required us to use the same calibrators hence the same 
diameter estimates.  The errors on the visibilities are therefore not 
independent.  It is the purpose of this paragraph to establish a 
method to compute this correlation and, more important, to be able to 
trace it to compute it {\it a posteriori} long after taking the data 
at the telescopes.\\
Let $S_{1}$ and $S_{2}$ be two spatial frequencies at which squared 
visibilities $V^{2}(S_{1})$ and $V^{2}(S_{2})$ have been measured.  
The visibility estimates of Eqs.~(\ref{eq:vis}) and 
(\ref{eq:visfinale}) can take the form:
\begin{equation}
\left\{
\begin{array}{ccc}
    V^{2}(S_{1}) & = & \alpha_{1}{V^{a}}^{2}(S_{1}) + 
    \beta_{1}{V^{b}}^{2}(S_{1}) \\
    V^{2}(S_{2}) & = & \alpha_{2}{V^{c}}^{2}(S_{2}) + 
    \beta_{2}{V^{d}}^{2}(S_{2})
\end{array}
\right.
\end{equation}
where the calibrators are A, B, C and D. To save room, the calibrator 
visibilities are replaced by the capital letters.  It can be shown 
that the correlation factor between the two squared visibilities is:
\begin{eqnarray}
    \label{eq:correlation}
\rho(V^{2}(S_{1}),V^{2}(S_{2})) & = & \frac{\alpha_{1}\alpha_{2}\sqrt{Var(A)Var(B)}\rho(A,C)}{\sqrt{Var(V^{2}(S_{1}))Var(V^{2}(S_{2}))}} \\ \nonumber 
& \times & \frac{\alpha_{1}\beta_{2}\sqrt{Var(A)Var(D)}\rho(A,D)}{\sqrt{Var(V^{2}(S_{1}))Var(V^{2}(S_{2}))}} \\ \nonumber
& \times & \frac{\beta_{1}\alpha_{2}\sqrt{Var(B)Var(C)}\rho(B,C)}{\sqrt{Var(V^{2}(S_{1}))Var(V^{2}(S_{2}))}} \\ \nonumber
& \times & \frac{\beta_{1}\beta_{2}\sqrt{Var(B)Var(D)}\rho(B,D)}{\sqrt{Var(V^{2}(S_{1}))Var(V^{2}(S_{2}))}}
\end{eqnarray}
When all calibrators are different, all correlations are zero. 
The correlation is maximum when a single calibrator has 
systematically been used. Measurement noise is present at the 
denominator only and the correlation is of course all the larger as 
the measurement noise is smaller.\\
The correlation between two expected squared visibilities at two 
different baselines is not easy to evaluate analytically.  Besides, it 
may depend upon the model of the calibrator.  A computation can be 
performed which shows that the correlation is indeed equal to 1 with 
an excellent accuracy as long as no baseline is equal to 0.  This can 
also be shown by expanding the visibility function.  Thus, the 
expected visibilities derived from a uniform disk model of a same 
calibrator at two different baselines are fully correlated to the 
first order.  This holds as long as the second derivative of the model 
is small (which in the case of the uniform disk model is true except 
close to the zeros of the model) and as a condition, none of the 
baselines is very close to zero.  In practice, the error on the 
diameter being usually small (less than 5\%), the first order 
approximation is valid and the two expected squared visibilities can 
therefore be considered fully correlated.  This is true down to very 
short baselines as for example for a diameter of $10\pm0.5\,$mas the 
correlation starts to decrease for a baseline below 5\,cm.  \\
For practical use, Eq.~(\ref{eq:correlation}) can be simplified as the 
correlations between expected visibilities are either 0 or 1 when the 
calibrators are respectively different or alike.  The only requirement 
to compute this correlation is therefore that the variances of the 
expected visibilities and the coefficients $\alpha$ and $\beta$ be 
saved with the reduced data.  These correlations will have to be 
computed to model fit the data.  The generalization of the 
Levenberg-Marquardt method with correlated data is given in the 
Appendix at the end of the paper.

\section{Validation of the method}
\label{sec:examples}
Examples of data reduction results and calibrations are presented. The 
quantities introduced in the previous sections are discussed in 
practical situations and general comments on observing strategies are 
expressed.
\subsection{Squared coherence factors statistics}
\label{sec:mu2stat}
I have plotted in Fig.~\ref{fig:histo} three examples of $\mu^{2}$ 
distributions.  In the case of V636~Her, the fringe speed puts the 
fringe frequency far above the turbulence piston spectrum.  The piston 
is almost frozen during each scan and the amount of correlated noise 
is small.  In the case of 71~UMa, the fringe speed is lower and the 
measurements are more sensitive to piston hence the higher correlated 
noise.  $\delta$~Sge is an intermediate case.  In all three examples, 
the distributions of $\mu^{2}$ are compatible with Gaussian 
distributions hence validating the basic assumption on the statistics 
of the $\mu^{2}$.  An important fact is that the amount of correlated 
noise is not negligible and must be taken into account.  However, a 
test on distributions is performed to detect deviations from Gaussian 
statistics.  Deviations are not common and are always due to 
instrumental problems.  In such cases, depending on the required level 
of data quality, data may be eliminated.

\subsection{Examples of transfer functions}
Figure~\ref{fig:coT2} presents two examples of squared co-transfer 
functions.  Full circles are measurements on calibrators whereas open 
circles are interpolations for science targets.  It is visible that 
the co-transfer function is not always stable and may experience 
variations.  In some cases like on May 15, 2000 at 8:07, an error of 
calibration may have happened as the co-transfer functions jump by a 
few percent.  Yet, in most cases, the transfer functions variations 
are slow on time scales of a few hours and variations can be well 
approximated to the first order. Data collected on May 22, 2000 show 
that this is still the case when the calibrator diameters are known 
with a very good precision.

\subsection{Discussion of model fitting and examples}
\subsubsection{Amount of correlation}
Before presenting examples let us summarize the different levels of 
correlations we have encountered so far:
\begin{enumerate}
    \item  correlation of coherence factors (coaxial beamcombiners)

    \item  correlation of interferometric channels (coaxial 
    beamcombiners)
    
    \item correlation of simultaneous baselines 

    \item  correlation of non-simultaneous baselines
\end{enumerate}
The first level ($r$) was addressed in Sect.~\ref{sec:mu2stat}.  
The amount of correlation between interferometric channels 
($\rho_{12}$) for a coaxial beamcombiner like FLUOR varies from a few 
percent for faint sources calibrated by very well-known calibrators 
to almost 100\% for bright sources calibrated by sources whose 
diameters are known with an accuracy of a few percent.  The two 
channels are therefore not fully independent in this case and it is 
important to check the $\chi^{2}$ defined by Eq.~(\ref{eq:chi2}).  
A large $\chi^{2}$ may indicate that either the assumptions on 
Gaussian statistics were wrong for these particular data or that the 
transfer function variation is not well measured.  In either case, 
data should be examined in detail to decide whether the visibility 
value can be used or not.  A blind method is to reject visibilities 
with a $\chi^{2}$ above a certain level that can be of 3 for difficult 
programs or relaxed to a larger value for easier programs.  It is 
important to note that if the transfer function has varied accordingly 
in both channels at the time the science target was observed by an 
amount larger than the error bars then this $\chi^{2}$ test will fail 
to detect it. It can only be detected if the variations are opposite 
in the two channels. This is certainly a weakness.\\
It will be interesting to assess the level of correlations of 
visibilities measured with multiple beam interferometers.  It can be 
anticipated that it will not be negligible and will be of the same 
level as $\rho_{12}$.  \\
The importance of correlation between visibilities recorded separately 
is illustrated in Fig.~\ref{fig:swvir}. The data have been reduced 
in two different ways. Data plotted with open circles and fitted by a 
dashed-line uniform disk model are reduced without taking 
correlations into account. Data plotted with full circles and fitted 
by the continuous line were reduced with the method of this paper. In 
the first case, the fit is of very good quality with a $\chi^{2}$ 
smaller than 1. Yet, all visibilities have been calibrated with the 
same source, hence a strong correlation between visibility values as 
the $3\times 3$ correlation matrix shows:
\begin{equation}
    C= \left[\begin{array}{ccc}
                   1 & 0.96 & 0.96  \\
                   0.96 & 1 & 0.97 \\
                   0.96 & 0.97 & 1
             \end{array}\right]
\end{equation}
It is to be noticed that the $\rho_{12}$ correlation factor is larger 
than 90\% for all three visibilities, a large fraction of this 
correlation being due to the common calibrator.  If correlations are 
ignored then noise is considered independent from one visibility to 
the other and this is why the first $\chi^{2}$ is smaller, as a large 
global noise is now interpreted as a large fluctuating noise from one 
visibility to the other.  On the contrary, when correlations are used, 
a tiny fraction of noise (4\% at most) can be considered a fluctuation 
giving degrees of freedom for the adjustement of the model.  This is 
equivalent to reducing error bars on visibilities by 96\% in the fit.  
The common noise due to the uncertainty on the calibrator is then a 
simple common bias on the visibilities but does not contribute to the 
noise in the fit, hence the much larger $\chi^{2}$.  In the zoomed 
part of this same figure, one can see that the fit now conforms to 
only one of the visibility data as the correlation matrix is close to 
being non-invertible (see the Appendix for the use of the correlation 
matrix in the fitting process).  In more physical terms, the 
correlations being very close to one, all data are equivalent and the 
fit can be derived from one of the visibility data.  If all data 
points were compatible despite the large correlations then the best 
fit curve would go through the error bars.  It is not the case here 
and this is why the $\chi^{2}$ is large.

\subsubsection{Examples of visibility accuracies}
Other examples of model fitting are presented in 
Fig.~\ref{fig:exemples}.  The BK~Vir data were calibrated with the 
same calibrator (the correlation matrix is similar to the matrix 
above) as SW~Vir but the visibilities are very consistent with each 
others.  Data for the three other sources are either totally 
independent or slightly correlated.  Only the first lobe data were 
used for the fit of G~Her.  These four examples show very good fits 
and consistency of data.  In particular, this sets the best absolute 
accuracy of the calibration of visibilities with FLUOR to 0.004 
(equivalent to an accuracy of 0.004 on $V^{2}$ with $V=0.5$ as 
$\sigma_{V^{2}}=2\sigma_{V}.V$ to the first order).

\subsubsection{Calibration strategies}
It is important to adapt the strategy of calibration to the type of 
astrophysical studies addressed with optical interferometers.  For 
most studies where visibility accuracies of a few percent are 
acceptable, the repeated use of a single or of a few calibrators is 
possible.  For difficult programs like exoplanet detection, a very 
high level of accuracy is required and the strategy needs to be well 
prepared.  Two cases may arise depending on whether the required 
calibration of visibilities is absolute or relative.  If absolute 
accuracies better than 0.001 have to be obtained on visibilities then 
it is very likely that no calibrator can be used twice, unless the 
error on the expected visibility of this calibrator is less than the 
level of accuracy required.  This would suppose that the visibility 
model of the calibrator be measured first.  Another possibility is 
relative detection.  As illustrated by the example of SW~Vir, if the 
same calibrator is systematically used, the fit is sensitive to very 
low levels as the correlated noise does not contribute to the value of 
the $\chi^{2}$.  In this example, a departure from the uniform disk 
model or a calibration error may have been detected to a level much 
lower than the error bars.  For very faint detail detection, this can 
work if the visibility curve of the calibrator is smooth and without 
wiggles of similar amplitude as the ones searched for on the science 
target.  \\
In any case, the observing strategy should be prepared in advance and 
should take the problem of data correlations into account.

\section{Conclusion}
I have proposed in this paper a method to calibrate visibility data 
obtained with single mode interferometers.  The single mode character 
is required to make valid the assumption that the statistics of 
coherence factors data are Gaussian and stationary.  It is possible to 
derive reliable error bars if all correlations are considered in the 
derivation of all estimators.  Correlations also need to be taken into 
account when fitting the data by models.  The validity of the method 
has been demonstrated on real interferometric data recorded with 
FLUOR. An important conclusion of this work is that the strategy of 
calibration has to be adapted for specific programs requiring high 
standards of calibration. 
%

\appendix

\section{Practical implementation of model fitting with correlated 
visibilities}
Algorithms for model fitting are well known. One of the most commonly 
used is the Levenberg-Marquardt algorithms. An example of practical 
implementation is given in \cite{press1988} in the case where data are 
not correlated. Here, I give a generalization of this algorithm to the 
case of correlated data.
\subsection{Definitions}
Let $(V_{i}^{2})_{i=1,\ldots,N}$ be a series of squared visibility 
measurements of an astronomical source. Let $C$ be the matrix of 
variances-covariances of these measurements:
\begin{equation}
    C= \left[\begin{array}{cccc}
                    \sigma_{1}^{2} & \rho_{1,2}\sigma_{1}\sigma_{2} & \cdots & \rho_{1,N}\sigma_{1}\sigma_{N}  \\
                    \rho_{1,2}\sigma_{1}\sigma_{2} & \sigma_{2}^{2} & 
                    \cdots & \vdots  \\
                    \vdots & \vdots & \ddots &  \vdots \\
                    \rho_{1,N}\sigma_{1}\sigma_{N} & \cdots & \cdots & \sigma_{N}^{2}
             \end{array}\right]
\end{equation}
Let $(M_{i}(\theta))_{i=1,\ldots,N}$ be the model to fit the data 
with $\theta$ a vector of $p$ parameters. It is assumed that the noise 
on the squared visibilities is Gaussian.

\subsection{Method}
The best estimates of the parameters in the sense of the highest 
likelihood are obtained by maximizing the likelihood function of the 
$Y=[V^{2}-M(\theta)]$ vector which leads to minimizing the following 
functional:
\begin{equation}
    S(\theta)=\frac{1}{2}Y^{t}C^{-1}Y
\end{equation}
The optimum $\hat{\theta}$ is obtained when:
\begin{equation}
    \left\{\begin{array}{ccc}
            \frac{\partial S}{\partial \theta} & = & 0  \\
            \frac{\partial^{2} S}{\partial \theta^{2}} & > & 0
        \end{array}\right.
\end{equation}
Besides, the covariance matrix of the estimated parameters is 
asymptotically ($N\rightarrow+\infty$) equal to:
\begin{equation}
    Var({\hat{\theta}}) = \left[\left.\frac{\partial^{2} S}{\partial 
        \theta^{2}}\right|_{\theta=\hat{\theta}}\right]^{-1}
\end{equation}
The search for the optimum can be performed with a generalized 
Levenberg-Marquardt algorithm.

\subsection{Generalized Levenberg-Marquardt algorithm}
Close to minimum, $S$ can be expanded to the second order in $\theta$:
\begin{equation}
    S(\theta)=S(\hat{\theta})+\left[\frac{\partial S}{\partial 
 \theta}\right]^{t}\left[\theta-\hat{\theta}\right]+
 \frac{1}{2}\left[\theta-\hat{\theta}\right]^{t}
 \left[\frac{\partial^{2} S}{\partial \theta^{2}}\right]
 \left[\theta-\hat{\theta}\right]
\end{equation}
The first and second derivative of S are respectively a vector and a 
matrix whose elements are:
\begin{eqnarray}
    \frac{\partial S}{\partial \theta_{k}} & = & - \left[\frac{\partial M}{\partial 
    \theta_{k}}\right]^{t}C^{-1}\left[V^{2}-M\right] \;\; , 
    k=1,\ldots,p \\ \nonumber
    \frac{\partial^{2} S}{\partial \theta_{k}\partial \theta_{l}} & = &  \left[\frac{\partial M}{\partial 
    \theta_{k}}\right]^{t}C^{-1}\frac{\partial M}{\partial \theta_{l}}
    -\left[ \frac{\partial^{2} M}{\partial \theta_{k}\partial \theta_{l}}\right]
    C^{-1} \left[V^{2}-M\right] \\ \nonumber
    & &   k,l=1,\ldots,p
\end{eqnarray}
In case $S$ is equal to its second order expansion then 
$\hat{\theta}$ can be guessed directly through:
\begin{equation}
    \hat{\theta}=\theta-\left[\frac{\partial^{2} S}{\partial 
    \theta^{2}}\right]^{-1}\left[\frac{\partial S}{\partial 
 \theta}\right]
\end{equation}
If the expansion is a poor approximation of $S$ then a steepest 
descent method has to be applied with:
\begin{equation}
    \theta_{\mathrm{next}}=\theta-\mathrm{constant}\times \left[\frac{\partial S}{\partial 
 \theta}\right]
\end{equation}
This the basis of the Levenberg-Marquardt method. It is classical to 
use the following notations:
\begin{eqnarray}
    \beta_{k} & = & - \frac{\partial S}{\partial 
 \theta_{k}} \;\; , k=1,\ldots,p \\ \nonumber
    \alpha_{kl} & = & \left[\frac{\partial M}{\partial 
     \theta_{k}}\right]^{t}C^{-1}\frac{\partial M}{\partial 
     \theta_{l}} \;\; , k,l=1,\ldots,p
\end{eqnarray}
where $\frac{\partial M}{\partial \theta_{k}}$ is the $k^{th}$ column 
of the matrix:
\begin{equation}
    \frac{\partial M}{\partial \theta}=\left[\frac{\partial 
    M(x_{i},\theta)}{\partial \theta_{k}}\right]_{i,k}
\end{equation}
The matrix equivalent of the above definition is therefore:
\begin{eqnarray}
    \beta & = & \left[\frac{\partial M}{\partial 
    \theta}\right]^{t}C^{-1}\left[V^{2}-M\right] \\ \nonumber
    \alpha & = & \left[\frac{\partial M}{\partial 
    \theta}\right]^{t}C^{-1}\frac{\partial M}{\partial 
    \theta}
\end{eqnarray}
At optimum the variance-covariance matrix of the parameters is:
\begin{equation}
    Var(\hat{\theta})=[\alpha]^{-1}
\end{equation}
With these definitions, $2S(\theta)$ is a $\chi^{2}$ with $N-p$ 
degrees of freedom. The increment in parameter space is therefore the 
solution of the linear system:
\begin{equation}
    \sum_{l=1}^{p}\alpha_{kl}\delta\theta_{l}=\beta_{k}
\end{equation}
when the quadratic form is a good approximation to $S$ otherwise the 
increment is given by:
\begin{equation}
    \delta\theta_{l}=\mathrm{constant}\times \beta_{l}
\end{equation}
The reader is referred to \cite{press1988} for the implementation of 
this modified algorithm.


\begin{thebibliography}{}
\bibitem[Chelli et al. (2000)]{chelli2002} Chelli, A., 2000, AMBER report AMB-IGR-017
\bibitem[Coud\'e du Foresto et al. (1997)]{foresto1997} Coud\'e du 
Foresto, 
V., Ridgway, S.T., Mariotti, J.-M., 1997, A\&ASS, 121, 379
\bibitem[Mourard et al. (1994)]{mourard1994} Mourard, D., 
Tallon-Bosc, I., Rigal, F., et al. 1994, A\&A 288, 675
\bibitem[Pelat (1992)]{pelat1992} Pelat, D., 1992, Cours ``Bruits et 
Signaux", \'Ecole Doctorale d'\^{I}le de France, 
Astronomie-Astrophysique
\bibitem[Perrin (1996)]{perrin1996} Perrin, G., 1996, PhD 
thesis, Universit\'e Paris VII 
\bibitem[Perrin et al. (1998)]{perrin1998} Perrin, G., Coud\'e du 
Foresto, V. Ridgway, S.T., et al., 1998, A\&A 331, 619
\bibitem[Perrin (2002)]{perrin2002} Perrin, G., 2002,  A\&A, in press
\bibitem[Press et al. (1988)]{press1988} Press, W.H., Flannery, B.P., 
Teukolsky, S.A., Vetterling, W.T., 1988, "Numerical Recipes in C", 
Cambridge University Press
\bibitem[Segransan et al. (1999)]{segransan2002} Segransan, D., 
Forveille, T., Millan-Gabet, R., Perrier, C., Traub, W.A., 1999, 
``Working on the Fringe: Optical and IR Interferometry from Ground and 
Space",  ASP Conference Vol.  194, p. 290,   S. Unwin and R. Stachnik 
Eds.


\end{thebibliography}
\end{document}